\documentclass[a4paper,fleqn,twocolumn,usenatbib]{mnras}

\usepackage{mathptmx}
\usepackage{ulem}
\usepackage[T1]{fontenc}
\usepackage{ae,aecompl}

\usepackage{graphicx}	% Including figure files
\usepackage{amsmath}	% Advanced maths commands
\usepackage{amssymb}	% Extra maths symbols
\usepackage{indentfirst}
\usepackage{bm}

\def\be{\begin{equation}}
\def\ee{\end{equation}}
\def\ba{\begin{eqnarray}}
\def\ea{\end{eqnarray}}
\usepackage{mathptmx}
\usepackage{indentfirst}
\usepackage{ulem}
\usepackage{graphicx}	% Including figure files
\usepackage{amsmath}	% Advanced maths commands
\usepackage{amssymb}	% Extra maths symbols
\usepackage{amsmath}
\usepackage{listings}
\usepackage{color}
\usepackage{url}

\graphicspath{{./}}

\title[Probing the Cosmological Principle with the CSST]{Probing the Cosmological Principle with the CSST Photometric Survey}

\author[Xu et al.]{
Yu-Tian Xu,$^{1}$
Ji-Ping Dai,$^{1, 2}$
Dong Zhao,$^{1}$
Jun-Qing Xia,$^{1}$\thanks{E-mail: xiajq@bnu.edu.cn}
\\
$^{1}$Department of Astronomy, Beijing Normal University, Beijing, 100875, China \\
$^{2}$Shenzhen Middle School, Shenzhen, 518024, China
}

\date{Accepted XXX. Received YYY; in original form ZZZ}

\pubyear{2022}

\begin{document}

\label{firstpage}
\pagerange{\pageref{firstpage}--\pageref{lastpage}}
\maketitle

\begin{abstract}
  The cosmological principle states that our Universe is statistically homogeneous and isotropic at large scales. However, due to the relative motion of the Solar System, an additional kinematic dipole can be detected in the distribution of galaxies, which should be consistent with the dipole observed in the cosmic microwave background (CMB) temperature. In this paper, we forecast the mock number count maps from the China Space Station Telescope photometric survey to reconstruct the kinematic dipole. Using the whole photometric mock data, we obtain a positive evidence for the dipole signal detection at $3\sigma$ confidence level, and the significance would be increased to $4\sigma$ when we only use the high-redshift samples with $z=1.8 \sim 4$. This result can provide a good consistency check between the kinematic dipoles measured in the CMB and that from the large scale structure, which can help us to verify the basic cosmological principle.
\end{abstract}

\begin{keywords}
cosmology: observations; cosmology: theory; large-scale structure of universe
\end{keywords}

\section{Introduction}
A fundamental hypothes of the current standard cosmological model is that our Universe is statistically homogeneous and isotropic at large scales. This is called cosmological principle and it applies for the cosmic rest-frame.
Such a hypothesis would promote the development of the modern cosmological model. Due to the motion of the Solar System relative to the cosmic rest-frame, the CMB temperature has an additional kinematic dipole. Thus, our relative velocity can be measured with high accuracy by the dipole anisotropy in the CMB temperature, which corresponds to $v=369.82 \pm 0.11 \mathrm{~km} / \mathrm{s}$ towards the direction $(l, b) = (264.021 \pm 0.011^ {\circ}, 48.253 \pm 0.005^ {\circ} )$ in galactic coordinates \citep{Planck:2018nkj}.

In the past few decades, the kinematic dipole in the distribution of extragalactic sources has been checked independently. It was firstly proposed by \citet{1984MNRAS.206..377E} and tested by \citet{Baleisis:1997wx}  using the Green Bank 1987 and Parkes-MIT-NRAO catalogues. \citet{Blake:2002gx} used the radio sources from NRAO VLA Sky Survey (NVSS), which has five times more samples than that used in \citet{Baleisis:1997wx}'s work, and they found a detection of our relative motion at more than 2$\sigma$ confidence level (C.L.). The result was consistent with the CMB dipole within $2\sigma$ C.L.  for both amplitude and direction. Later, the analyses of NVSS radio galaxies are revisited in many other works \citep{Singal:2011dy, Rubart:2013tx, Tiwari:2013vff, Fernandez-Cobos:2013fda, Tiwari:2013ima, Bengaly:2017slg}, and they found consistent dipole directions with the one in the CMB but the dipole amplitudes were larger than that seen in the CMB. There are also some explanations for this large dipole amplitudes, such as the contamination from local structure \citep{Colin:2017juj, Tiwari:2015tba}, or the  existence of a large void \citep{Rubart:2014lia}. Recent work showed the dipole amplitude is consistent with the CMB result using the extragalactic radio sources from the Very Large Array Sky Survey (VLASS) combined with the Rapid Australian Square Kilometer Array Pathfinder Continuum Survey (RACS) \citep{Darling:2022jxt}. \citet{Siewert:2020krp} analysed the TIFR GMRT Sky Surveys (TGSS), Westerbork Northern Sky Survey (WENSS), Sydney University Molonglo Sky Survey (SUMSS) and NVSS radio source catalogues, and they found an increasing dipole amplitude with decreasing frequency. 
Except for the radio sources, there are also some works probing the kinematic dipole by visible and infrared catalogues \citep{Itoh:2009vc, Gibelyou:2012ri, Yoon:2014daa, Alonso:2014xca, Javanmardi:2016whx, Rameez:2017euv}, and they did not find significant evidence for anomalous dipole. Recently, \citet{2021ApJ...908L..51S} used 1.36 million quasars observed by the Wide-field Infrared Survey Explorer (WISE) to study the kinematic dipole and they found its amplitude is over twice as large as expected.
\citet{Yoon:2015lta} forecasted  that an optical survey covering $\sim 75\%$ of the sky in both hemispheres and having $\sim 30$ million galaxies can detect the kinematic dipole at $5\sigma$ C.L., while its median redshift should be at least $z \sim 0.75$ for negligible bias from the local structure.

In this work, we focus on the China Space Station Telescope (CSST) photometric survey to reconstruct the kinematic dipole direction and amplitude. The CSST is a 2-meter space telescope in the same orbit of the China Manned Space Station \citep{Lin:2022aro}. The CSST survey will cover $17,500$ $\rm deg ^ {2} $ sky area with the field of view $1.1$ $\rm deg ^ {2} $. It will carry multiple scientific equipments, that allow it to collect photometric images and spectroscopic data in the meantime. It has seven photometric and three spectroscopic bands from near-UV to near-IR covering 255-1000 nm. The CSST photometric survey can observe billions of galaxies in the redshift range of $z=0 \sim 4$, which is expected to give a tight constraint on the kinematic dipole in the future.

This paper is organized as follows: In section \ref{sec:2}, we generate the mock CSST number count maps and include the effect of the fiducial kinematic dipole. Section \ref{sec:3} introduces the estimators used in this paper and exhibits the reconstruction results. Finally, we summarize the conclusions and give relevant discussions in Section \ref{sec:4}.

\section{Mock Data}
\label{sec:2}

\subsection{The angular power spectrum of galaxies}
To obtain the number count maps of the CSST photometric survey, we first need to compute the theoretical angular power spectrum of the survey.  The observed galaxy density contrast in a given direction $\hat{\boldsymbol{n}}_{1}$ is
\be
\delta_{\rm g}\left(\hat{\boldsymbol{n}}_{1}\right) =\int b_{\rm g}(z) n(z) \delta_{\rm m}\left(\hat{\boldsymbol{n}}_{1}, z\right) {\rm d} z,
\ee
where $b_{\rm g}(z)$ is the bias factor relating the galaxy overdensity to the mass overdensity, $\delta_{\rm g}  = b_{\rm g}\delta_{\rm m}$, $n(z)$  is the normalized windows function of the survey. So the angular power spectrum can be easily expressed in the harmonic space:
\be
\begin{aligned}
\label{eq:aps}
\left\langle\delta_{\rm g}\left(\hat{\boldsymbol{n}}_{1}\right) \delta_{\rm g}\left(\hat{\boldsymbol{n}}_{2}\right)\right\rangle &= \int b_{\rm g}(z) n(z) D(z) {\rm d} z \int b_{\rm g}\left(z^{\prime}\right)n\left(z^{\prime}\right) D\left(z^{\prime}\right) {\rm d} z^{\prime} \\
& \times \int \frac{{\rm d} k k^{2}}{(2 \pi)^{3}} P_{\rm m}(k,z=0) \int {\rm d} \Omega_{k} e^{{\rm i} \boldsymbol{k} \cdot \hat{\boldsymbol{n}}_{1} \eta_{1}} e^{-{\rm i} \boldsymbol{k} \cdot \hat{\boldsymbol{n}}_{2} \eta^{\prime}}.
\end{aligned}
\ee
Here, we used
\be
\left\langle\delta_{\mathrm{m}}(\boldsymbol{k}) \delta_{\mathrm{m}}\left(\boldsymbol{k}^{\prime}\right)\right\rangle=(2 \pi)^{3} \delta_{\mathrm{D}}\left(\boldsymbol{k}+\boldsymbol{k}^{\prime}\right) P_{\mathrm{m}}(k).
\ee
$D(z)$ is the growth factor, $P_{\rm m}(k)$ is the matter power spectrum, and $\eta$ is the comoving distance. Expanding out the exponentials yields:
\be
\int {\rm d} \Omega_{k} {\rm e}^{{\rm i} \boldsymbol{k} \cdot \hat{\boldsymbol{n}}_{1} \eta_{1}}  {\rm e}^{-{\rm i} \boldsymbol{k} \cdot \hat{\boldsymbol{n}}_{2} \eta^{\prime}} = 4 \pi \sum_{\ell}(2 \ell+1) j_{\ell}(k \eta) j_{\ell}\left(k \eta^{\prime}\right) P_{\ell}\left(\hat{\boldsymbol{n}}_{1} \cdot \hat{\boldsymbol{n}}_{2}\right),
\ee
where $j_{\ell}$ are the spherical Bessel functions and $P_{\ell}\left(\hat{\boldsymbol{n}}_{1} \cdot \hat{\boldsymbol{n}}_{2}\right)$ are the  Legendre functions. Thus we can rewrite Eq.(\ref{eq:aps}) as
\be
\left\langle\delta_{\rm g}\left(\hat{\boldsymbol{n}}_{1}\right) \delta_{\rm g}\left(\hat{\boldsymbol{n}}_{2}\right)\right\rangle =\sum_{\ell} \frac{(2 \ell+1)}{4 \pi} C_{\ell}^{\rm gg} P_{\ell}\left(\hat{\boldsymbol{n}}_{1} \cdot \hat{\boldsymbol{n}}_{2}\right),
\ee
and the angular power spectra $C_{\ell}^{\rm gg}$ is given by
\be
\label{eq:aclactual}
C_{\ell}^{\rm gg}=\frac{2}{\pi} \int k^{2} {\rm d} k P_{\rm m}(k,z=0)\left[I_{\ell}^{\rm g}(k)\right]^{2},
\ee
where $I_{\ell}^{\rm g}(k)$ are
\be
I_{\ell}^{\rm g}(k)=\int b_{\rm g}(z)n(z) D(z) j_{\ell}[k \eta(z)] {\rm d} z.
\ee

\subsection{The generation of mock data}
\label{sec:md}
\begin{figure}
  \centering
  \includegraphics[width=1.0\linewidth]{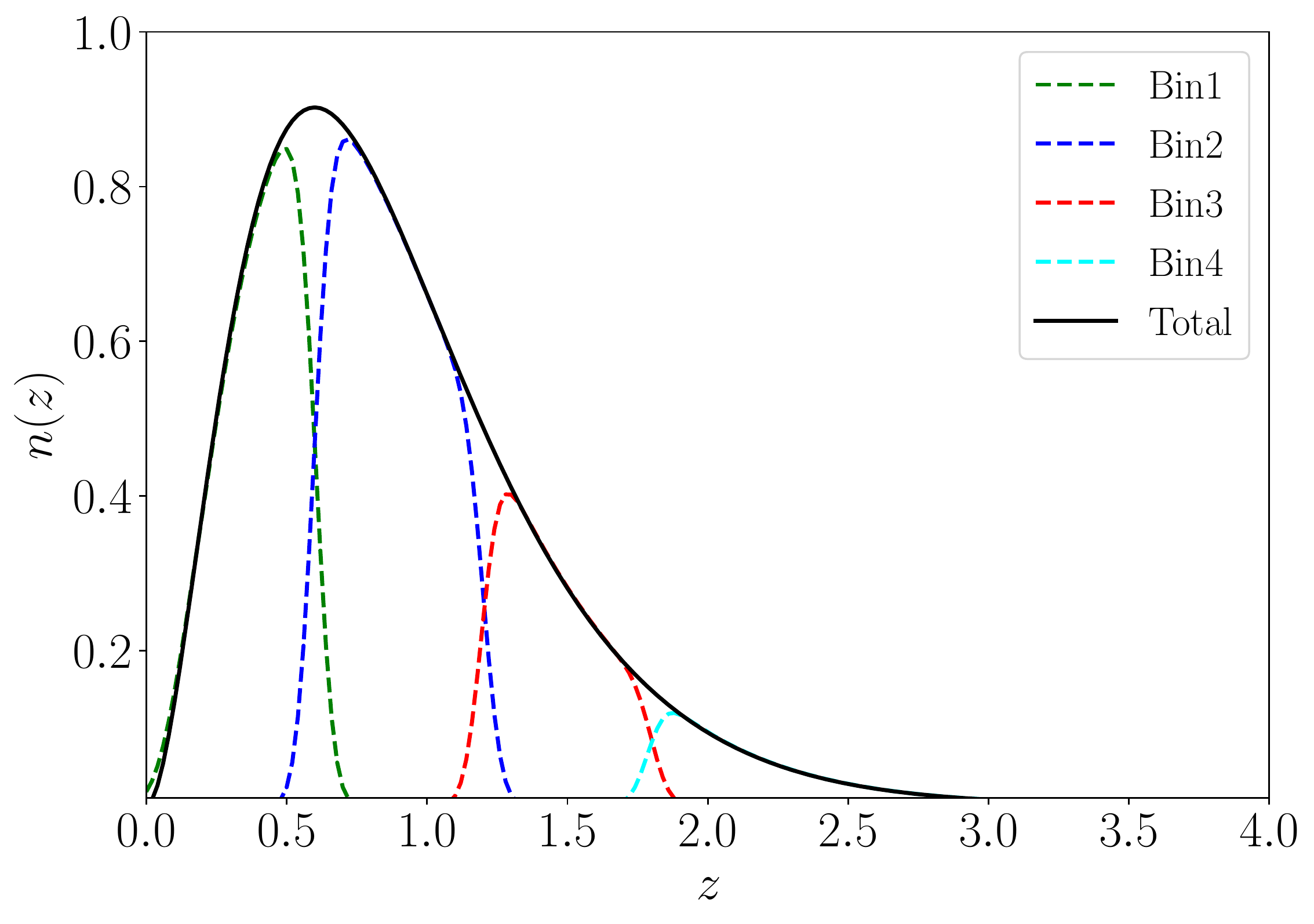}
  \caption{The selection function of the CSST photometric survey. The black solid curve shows the total redshift distribution, and the colored dotted curves show the $n_i(z)$ for the four tomographic bins.}
  \label{fig:dndz}
\end{figure}

According to Eq.(\ref{eq:aclactual}), we can see the main differences in the theoretical angular power spectrum come from the different selection functions and bias factors, so we have to explicit  these parameters to generate the mock data of  the CSST photometric
survey. \citet{2018MNRAS.480.2178C} have studied the selection function of the CSST photometric survey based on the COSMOS catalog \citep{2007ApJS..172...99C,2009ApJ...690.1236I}. They found the selection function has a peak around $z = 0.6$, and the range can extend to $z \sim 4$. For analytical use, we simply adopt a smooth function proposed by \citet{Lin:2022aro}, which takes the form of
\be
n(z) \propto z^2{\rm e}^{-z/z^{\star}},
\ee
where $z^{\star}=0.3$.
In Figure \ref{fig:dndz}, we show the normalized $n(z)$ in black solid curve.

We divide the redshift distribution into four tomographic bins to explore the differences between the reconstructed kinematic dipoles using different redshift ranges. The lower and upper limits of the four tomographic bins are: [0, 0.6], [0.6, 1.2], [1,2, 1.8], [1.8, 4], respectively. The real galaxy selection function in the $i$th tomographic bin can
be expressed as \citep{Gong:2019yxt}
\be
\label{eq:dndz}
n_{i}(z)=\int_{z_{\mathrm{p}, \mathrm{l}}^{i}}^{z_{\mathrm{p}, \mathrm{u}}^{i}} d z_{\mathrm{p}} n(z_{\rm p}) p\left(z_{\mathrm{p}} \mid z\right),
\ee
where $z_{\mathrm{p}, \mathrm{l}}^{i}$ and $z_{\mathrm{p}, \mathrm{u}}^{i}$ are the lower and upper limits of the $i$th
tomographic bin, respectively. $p\left(z_{\mathrm{p}} \mid z\right)$ is the photo-z distribution function given the real redshift $z$. We assume it takes the form of
\be
p\left(z_{\mathrm{p}} \mid z\right)=\frac{1}{\sqrt{2 \pi} \sigma_{z}} \exp \left[-\frac{\left(z-z_{\mathrm{p}}-\Delta z\right)^{2}}{2 \sigma_{z}^{2}}\right],
\ee
where $\Delta_z$ and $\sigma_z$ are the redshift bias and scatter, respectively. In our work, we assume $\Delta_z=0$ and $\sigma_z=0.05$, which are also used in \citet{Gong:2019yxt} and  \citet{Lin:2022aro}.

\begin{figure}
  \centering
  \includegraphics[width=1.0\linewidth]{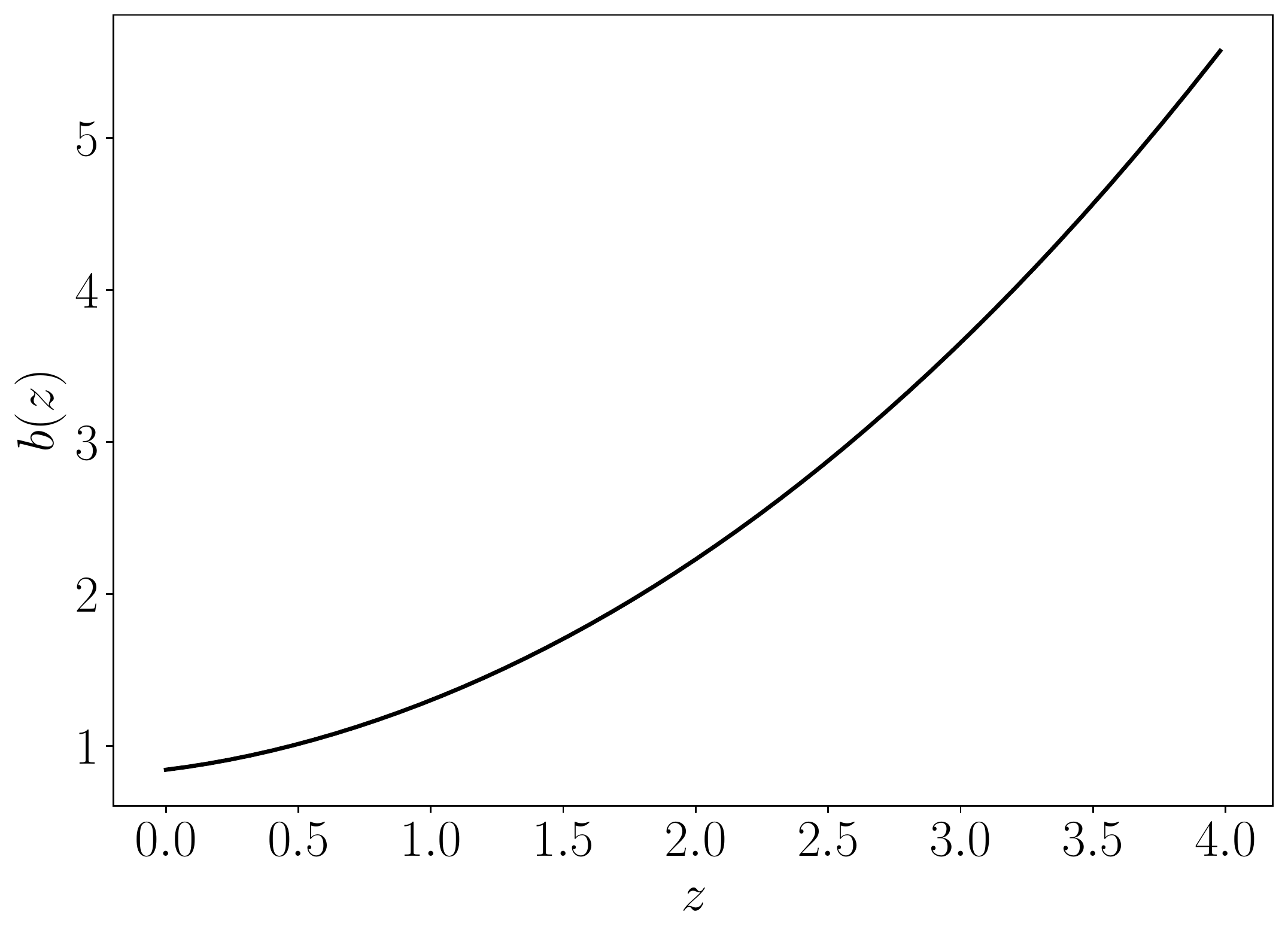}
  \caption{The galaxy bias as a function of redshift used in our work.}
  \label{fig:bias}
\end{figure}

As for the bias factor, we assume galaxies form in dark matter halos, so we can express the galaxy bias using the halo model, at least at large scales. We use the the fitting formula from \cite{Tinker:2010my} in our work, and choose the minimal halo mass as $M = 10^{12}~h^{-1}M_{\odot}$. The resulting bias factor is shown in Figure \ref{fig:bias}.

\begin{figure}
  \centering
  \includegraphics[width=1.0\linewidth]{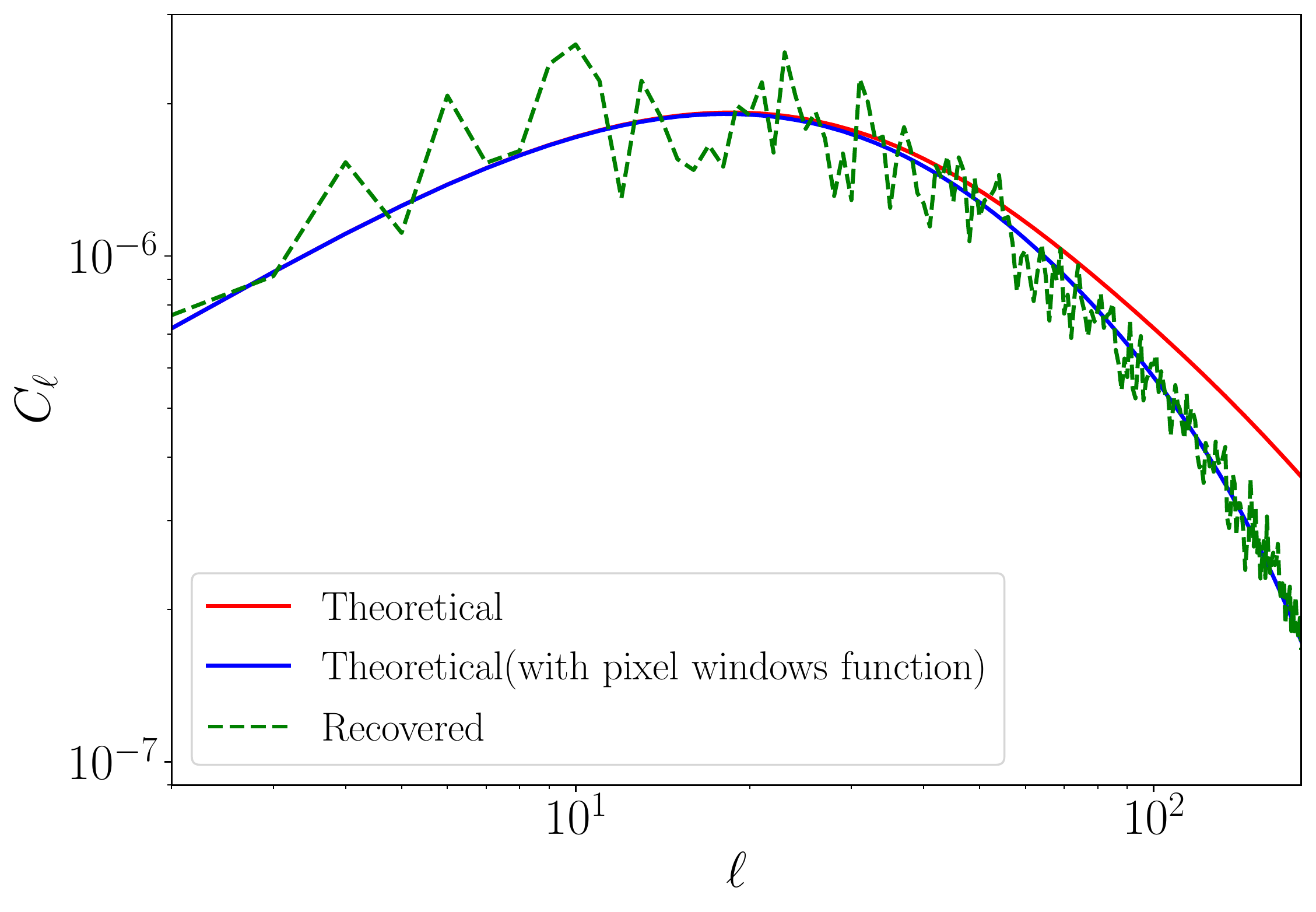}
  \caption{The angular power spectrum obtained from the full redshift distribution. The red and blue solid curve show the theoretical power spectra without and with considering the pixel windows function. And the green dotted line shows the recovered result from the mock data. Here, the other cosmological parameters are set to Planck's best fit results \citep{Aghanim:2018eyx}.}
  \label{fig:cla}
\end{figure}

Using the selection function and bias factor mentioned above, we compute the theoretical angular power spectrum of the CSST photometric survey and show the results in Figure \ref{fig:cla}. The red solid curve shows the angular power spectrum obtained from the full redshift distribution.

Next, we need to explicit the survey area of the CSST photometric survey. It will observe the sky where the latitude of the zodiac $|\beta|>20^{\circ}$, and we also mask the low galactic latitudes ($|b|<20^{\circ}$) to remove Milky Way objects. This amounts to effective sky area of $f_{\rm sky}\simeq 41\%$. The CSST photometric survey is expected to observe billions of galaxies in the the redshift range of $z=0\sim 4$, and the surface galaxy density of the four tomographic bins are 7.9, 11.5, 4.6, 3.7 $\rm arcmin^{-2}$ \citep{Gong:2019yxt}, respectively, therefore, the shot noise is not significant affect the angular power spectrum at large scales. Spatially varying dust extinction or instrumentation effects may arise additional systematic noise $C_{\rm sys}$, we follow \citet{Gong:2019yxt}'s work and assume $C_{\rm sys}=10^{-8}$, which is independent to tomographic bins or scales. From Figure \ref{fig:cla}, we can see the systematic noise is insignificant at $\ell<200$.

\begin{figure}
  \centering
  \includegraphics[width=1.0\linewidth]{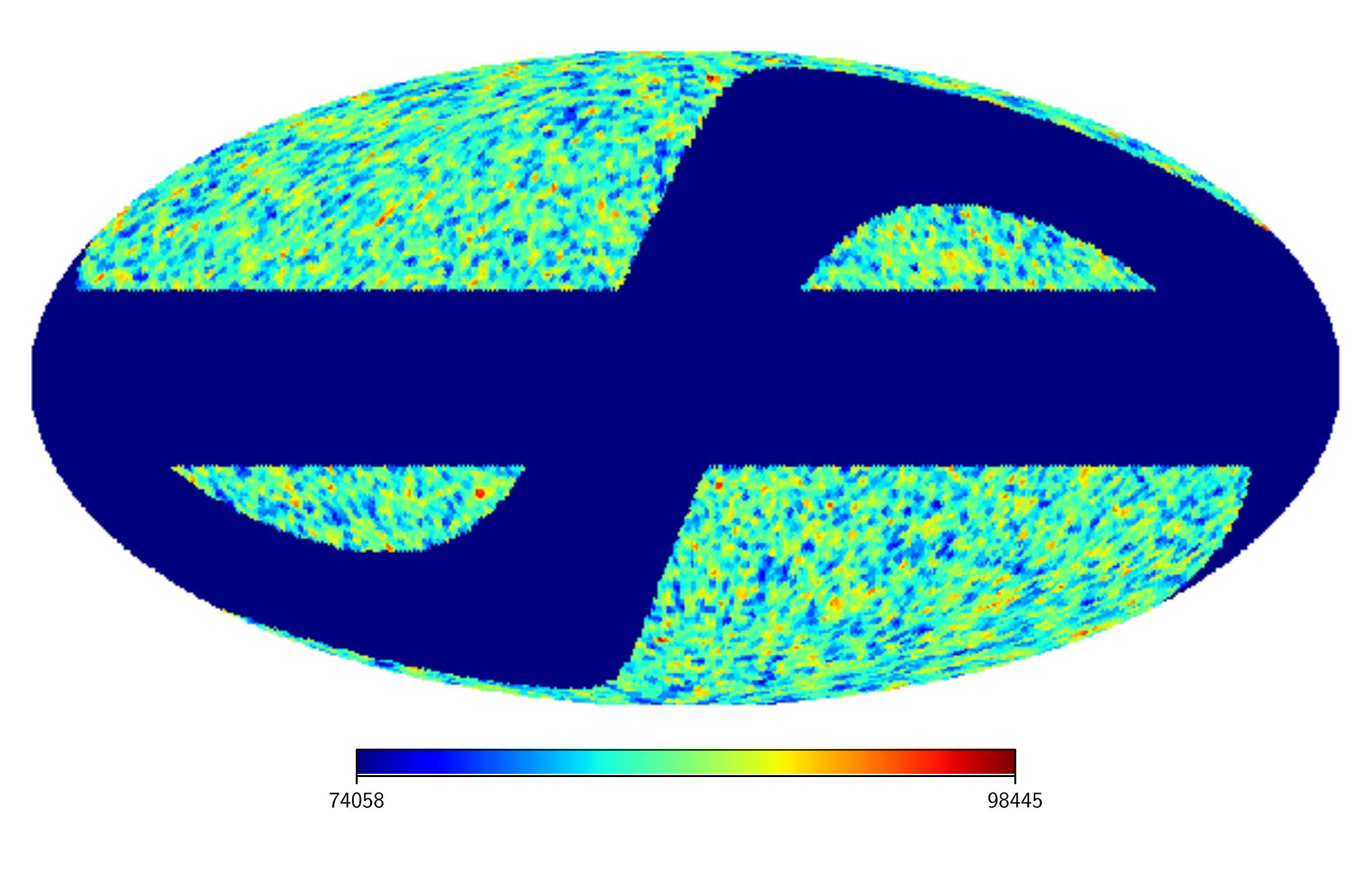}
  \caption{The simulated CSST number count map of the full redshift samples.}
  \label{fig:mapall}
\end{figure}

Finally, we input the theoretical angular power spectrum and the redshift distribution of galaxies to the public log-normal code FLASK \citep{Xavier:2016elr} to generate mock CSST number count maps. We use $N_{\rm side}$ = 64 for computational efficiency. Since we are mainly concerned with large scale information, this choice is sufficient for our calculation. The simulated number count map of the full redshift samples is shown in Figure \ref{fig:mapall}.

We can also compute the angular power spectrum from the mock data to ensure the accuracy of our calculation. The recovered spectrum is shown in Figure \ref{fig:cla} with the green dotted line. It is depressed at small scales due to the pixel windows function. If we include this effect in our theoretical spectrum, the recovered spectrum is well matched with the theoretical spectrum.

\subsection{Including the effect of the fiducial kinematic dipole}
\label{sec:dio}

We include the effect of the kinematic dipole through a dipole modulation of the number counts. The modified source counts can be expressed as \citep{Bengaly:2017slg, Bengaly:2018ykb}:
\be
\label{eq:Nobs}
N_{\text {obs }}(\hat{\boldsymbol{n_i}})=N_{\text {rest }}(\hat{\boldsymbol{n_i}})[1+A \hat{\boldsymbol{n_i}} \cdot \boldsymbol{\hat{\beta}}],
\ee
where $\hat{\boldsymbol{n_i}}$ is the direction of $i$th HEALPix cell in the map, $\boldsymbol{\hat{\beta}}$ is the direction of the kinematic dipole observed by Planck, which is $(l,b) \simeq (264^{\circ}, 48^{\circ})$ \citep{Planck:2018nkj}, and $A$ is the amplitude of kinematic dipole signal amplitude which is given by \citep{Burles:2006xf, Itoh:2009vc, Yoon:2015lta}
\be
\label{eq:A}
A = 2[1+1.25x(1-p)]\beta,
\ee
where $\beta = v/c \simeq 0.00123$  is measured with high accuracy by the Planck collaboration \citep{Planck:2018nkj}. The contribution $2\beta$ comes from relativistic aberration. The correction $[1+1.25x(1-p)]$ corresponds to the Doppler effect, where $x$ is the faint-end slope of the source counts, $x \equiv {\rm d}\log_{10}[n(m<m_{\rm lim})]/{\rm d}m_{\rm lim}$, and $p$ is the logarithmic slope of the intrinsic flux density power-law, $S_{\rm rest}(\nu)\propto \nu^{p}$.
\citet{Marchesini:2012yh} found that the faint end of the V-band galaxy luminosity function does not vary much and is equal to $ x = 0.11 \pm 0.02 $.
We use the mean value of 0.11 in our work.
Moreover, for optical sources, the flux density slope $p$ varies significantly with the age of the source. For simplicity, we choose $p=0$ as  \citet{Yoon:2015lta} used. Finally, Applying all these values to Eq.(\ref{eq:A}), we get the amplitude of kinematic dipole signal amplitude
\be
\label{eq:Av}
A\simeq0.0028.
\ee

Using Eq.(\ref{eq:Nobs}) and Eq.(\ref{eq:Av}), we can include the kinematic dipole to mock number count maps by a dipole modulation in pixelized source counts. \citet{Bengaly:2018ykb} also individually modulated each source according to the fiducial dipole signal, and they found these two methods lead to negligible differences in the dipole reconstruction.

\section{Reconstruction Results}
\label{sec:3}

\subsection{Estimator}

\begin{figure}
  \centering
  \includegraphics[width=1.0\linewidth]{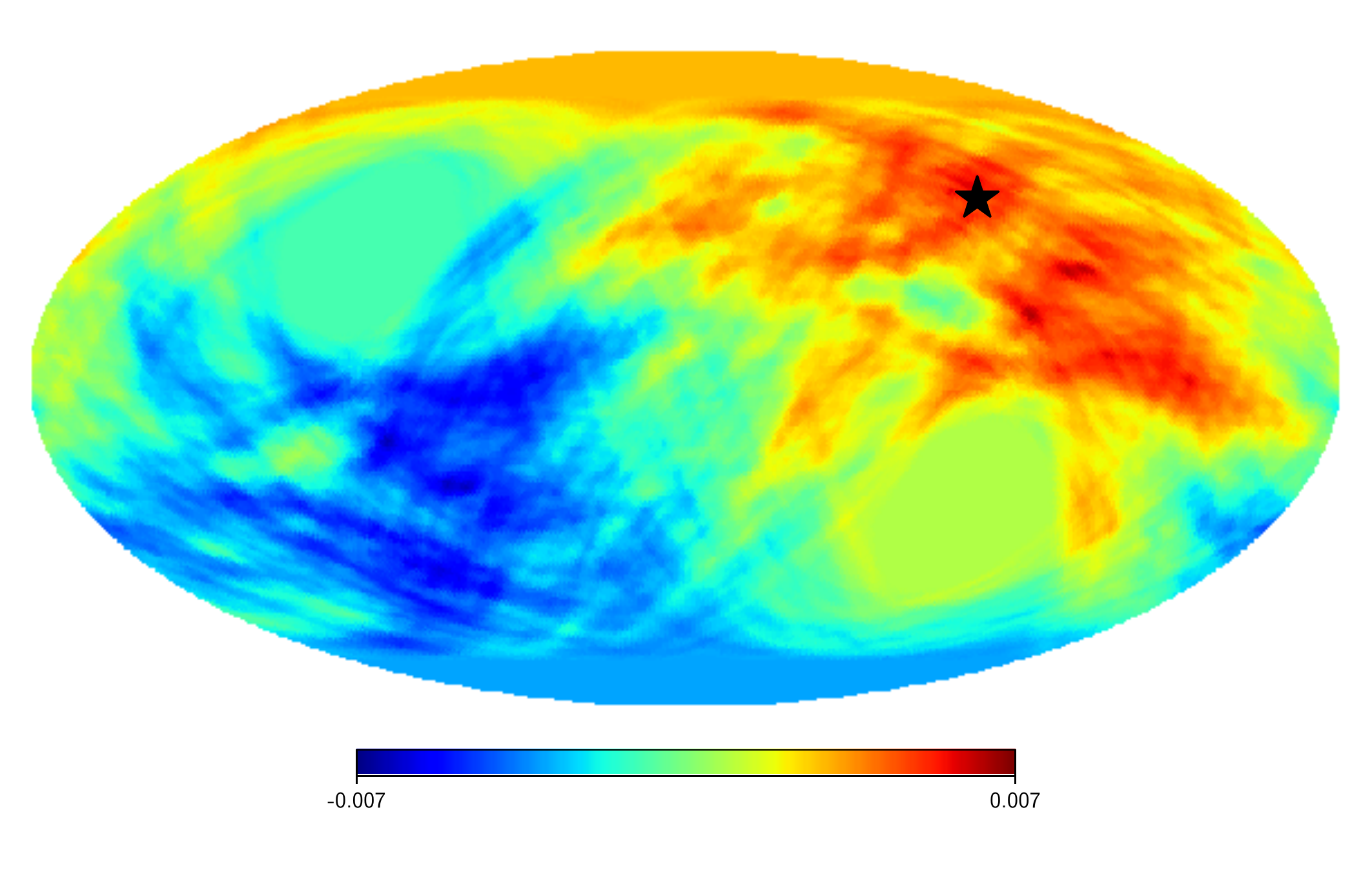}
  \caption{The $\Delta$-map from a mock data after including the kinematic dipole, and the black star indicates the direction of the fiducial kinematic dipole.}
  \label{fig:deltamap}
\end{figure}

Following \citet{Bengaly:2018ykb}, we reconstruct the kinematic dipole using both the linear estimator and the quadratic estimator. When considering the linear estimator, we can decompose the sky into 49152 HEALPix cells ($N_{\rm side}$ = 64), and counting the difference in the number of sources contained in opposite hemispheres whose symmetry axes are provided by these cell centres. Therefore, we can construct a $\Delta$-map, defined by
\be
\label{eq:deltamap}
\Delta(\hat{\boldsymbol{n_i}}) \equiv \frac{\sigma^{\rm U}(\hat{\boldsymbol{n_i}}) - \sigma^{\rm D}(\hat{\boldsymbol{n_i}})}{\sigma} = A \cos(\theta),
\ee
where $\hat{\boldsymbol{n_i}}$ is the direction of $i$th HEALPix cell,
$\sigma^{\rm U}(\hat{\boldsymbol{n_i}})$ and $\sigma_i^{\rm D}(\hat{\boldsymbol{n_i}})$ are the galaxy density of the `up' and `down' hemispheres divided by the axis direction $\hat{\boldsymbol{n_i}}$, $\sigma$ is the galaxy density of the whole map, $\theta$ is the angle between the $i$th cell centre and the  kinematic dipole  direction. In Figure \ref{fig:deltamap} we show the $\Delta$-map from a mock data after including the kinematic dipole, and the black star indicates the direction of the fiducial kinematic dipole.
From Figure \ref{fig:deltamap},
we can see when $\hat{\boldsymbol{n_i}}$ is closer to $\hat{\beta}$, the value of $\Delta(\hat{\boldsymbol{n_i}})$ is larger. The result can also be verified from Eq.(\ref{eq:deltamap}). Therefore, we can regard the maximum $\Delta(\hat{\boldsymbol{n_i}})$ value and its corresponding $\hat{\boldsymbol{n_i}}$ as the reconstructed kinematic dipole amplitude and direction, respectively.

\citet{Rubart:2013tx} pointed the linear estimator suffers from bias, so we also adopt a quadratic estimator. The main idea is to vary the dipole amplitude and direction to find the minimum of
\be
\sum_{i} \frac{\left[N_{\rm obs}(\hat{\boldsymbol{n_i}})-\bar{N}\left(1+\tilde{A} \cos \tilde{\theta}\right)\right]^{2}}{\bar{N}\left(1+\tilde{A} \cos \tilde{\theta}\right)}.
\ee
The sum is taken over all unmasked pixels, and $\bar{N}$ is the average number density of the whole map. $\tilde{A}$ and $\tilde{\theta}$ are the parameters to vary. In our work, we use a two-dimensional grid of dimensions with $N_{\tilde{\theta}}\times N_{\tilde{A}}$ as our parameter space. The dipole direction is probed along the pixel centres, so $N_{\tilde{\theta}} = 49152$. For the dipole amplitudes, we choose the range $[0.0005, 0.006]$ with 20 bins uniformly spaced in it, which means $N_{\tilde{A}}=20$.

\subsection{The reconstruction results using the full redshift samples}

\begin{figure}
  \centering
  \includegraphics[width=1.0\linewidth]{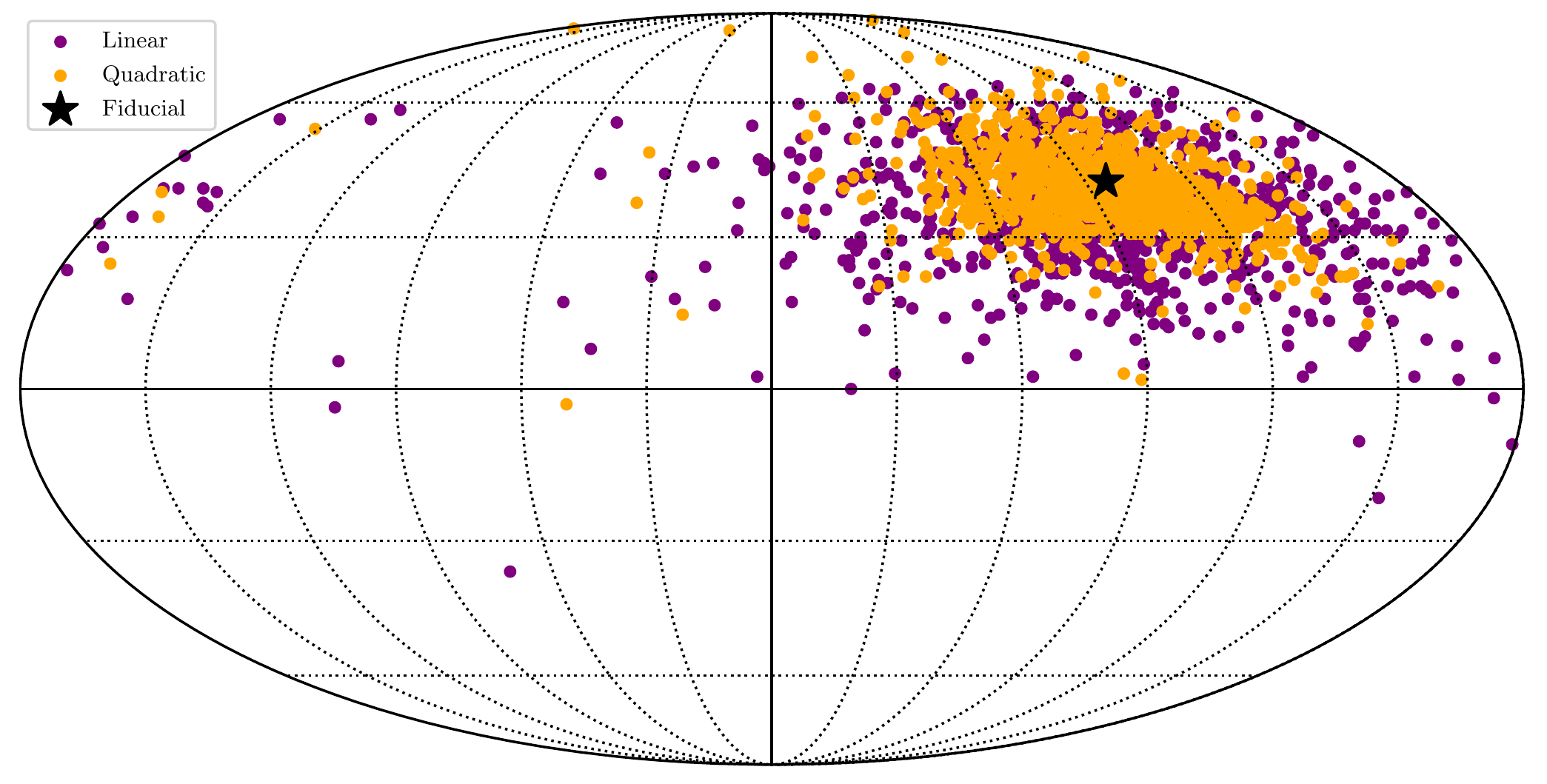}
  \caption{The reconstructed dipole directions using the  full redshift samples.  The purple and orange dots are obtained from the linear and quadratic estimators using 1000 simulations, and the black star is the fiducial value.}
  \label{fig:zall1}
\end{figure}

\begin{figure}
  \centering
  \includegraphics[width=1.0\linewidth]{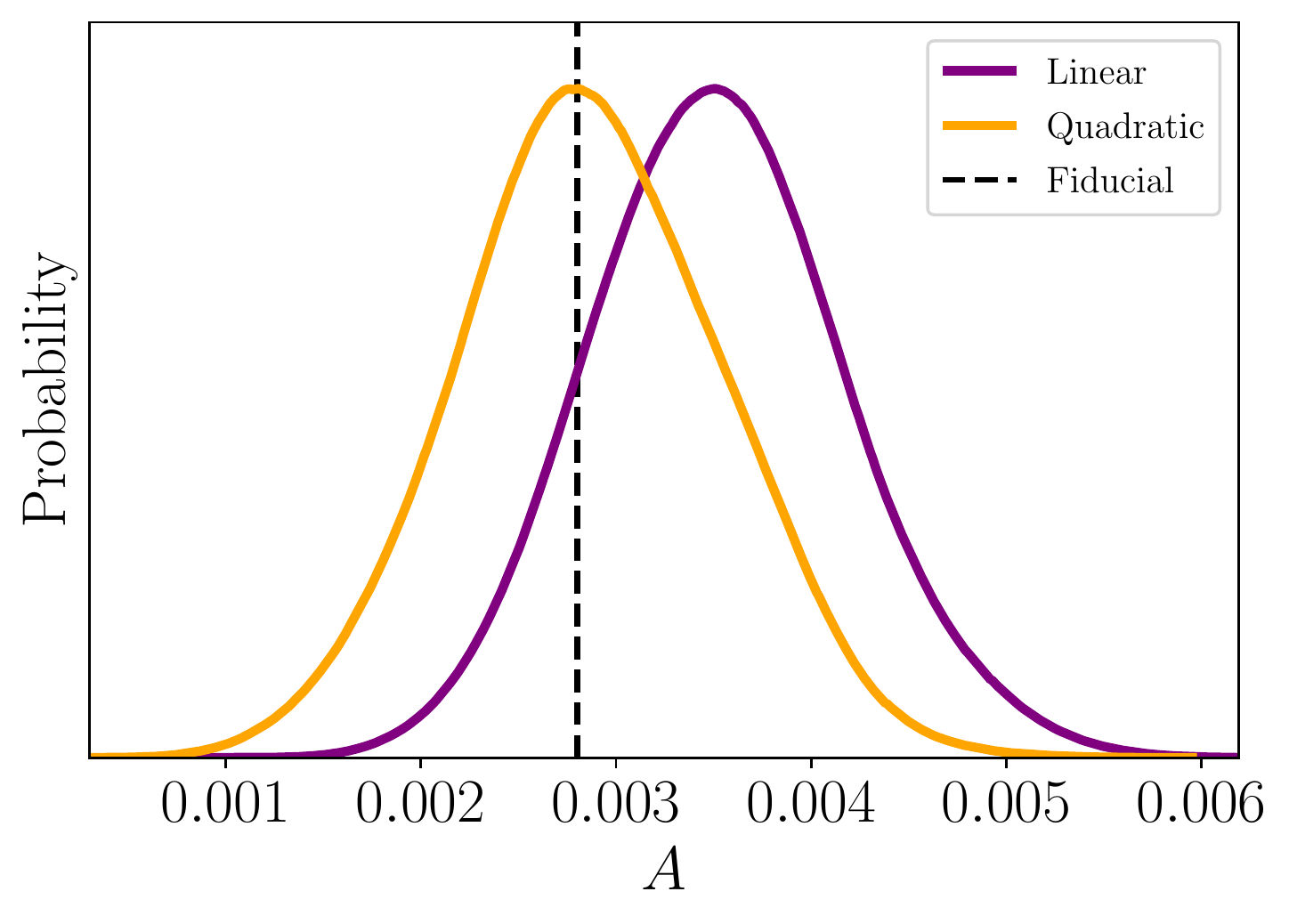}
  \caption{The one-dimensional distributions of reconstructed dipole amplitudes using the full redshift samples. The purple and orange lines are obtained from the linear and quadratic estimators, and the black dotted line is the fiducial value.}
  \label{fig:zall2}
\end{figure}

To obtain the statistical results, we simulate 1000 mock maps with the effect of the kinematic dipole using Eq.(\ref{eq:Nobs}). We use both the linear and quadratic estimators to reconstruct their dipole directions and amplitudes, and the results are shown in Figures \ref{fig:zall1} and \ref{fig:zall2}. For the latitude $b$ and the amplitude $A$, we can directly calculate their mean values and standard deviations from the 1000 simulations. But for the longitude $l$, we have to account for the fact that close to the pole, a small shift can lead to a large difference in $l$. We therefore use a weighted mean and standard deviation \citep{Bengaly:2018ykb}:
\be
\bar l = \frac{1}{n}\sum_{i}w_i l_i;~~~ \sigma_l^2 = \frac{1}{n-1}\sum_i w_i(l_i-\bar{l})^2,
\ee
where $n=1000$ is the number of mock maps, and $w_i$ are weights,
\be
w_i = \frac{n\sin b_i}{\sum_i \sin b_i}.
\ee

We list the quantitative results in Table \ref{tab:zall}. We can see the reconstructed dipole direction obtained from the linear estimator is $(l,b) = (259.2\pm50.3^{\circ}, 55.6\pm12.7^{\circ})$, and the result is $(l,b) = (265.1\pm30.8^{\circ}, 51.1\pm8.5^{\circ})$ using the quadratic estimator. These two results are consistent with the fiducial value, but the errors from the quadratic estimator are smaller than the ones from the linear estimator. For the dipole amplitude, the linear estimator gives a result larger than the fiducial value: $A = 0.00352 \pm 0.00085$, which is also verified by \citet{Rubart:2013tx}. This bias can be eliminated using the quadratic estimator. The reconstructed result is $A = 0.00296 \pm 0.00092$, which means we can detect the kinematic dipole signal at 3$\sigma$ C.L. using CSST photometric survey.

\begin{table}
 \caption{The reconstructed dipole directions and amplitudes using the full redshift samples. Here we list the results obtained from both the linear and quadratic estimators.}
 \begin{tabular}{c c c c}
    \hline
    \hline
    Method & $b(^{\circ})$ & $l(^{\circ})$ & $A$($\times 10^{-3}$) \\
    \hline
    Linear estimator & $55.6\pm12.7$ & $259.2\pm50.3$ & $3.52 \pm 0.85$\\
    Quadratic estimator & $51.1\pm8.5$ & $265.1\pm30.8$ & $2.96 \pm 0.92$\\
    \hline
    Fiducial value & 48.0 & 264.0 & 2.8\\
    \hline
 \end{tabular}
 \centering
 \label{tab:zall}
\end{table}

Comparing the linear and quadratic estimators, we find the latter method is better in our analysis, so we only focus on the quadratic estimator in the following work. However, the linear estimator is less compute-intense and easier to implement. It can be used to obtain fast order-of-magnitude checks.

\subsection{The reconstruction results with different redshift bins}

\begin{figure}
  \centering
  \includegraphics[width=1.0\linewidth]{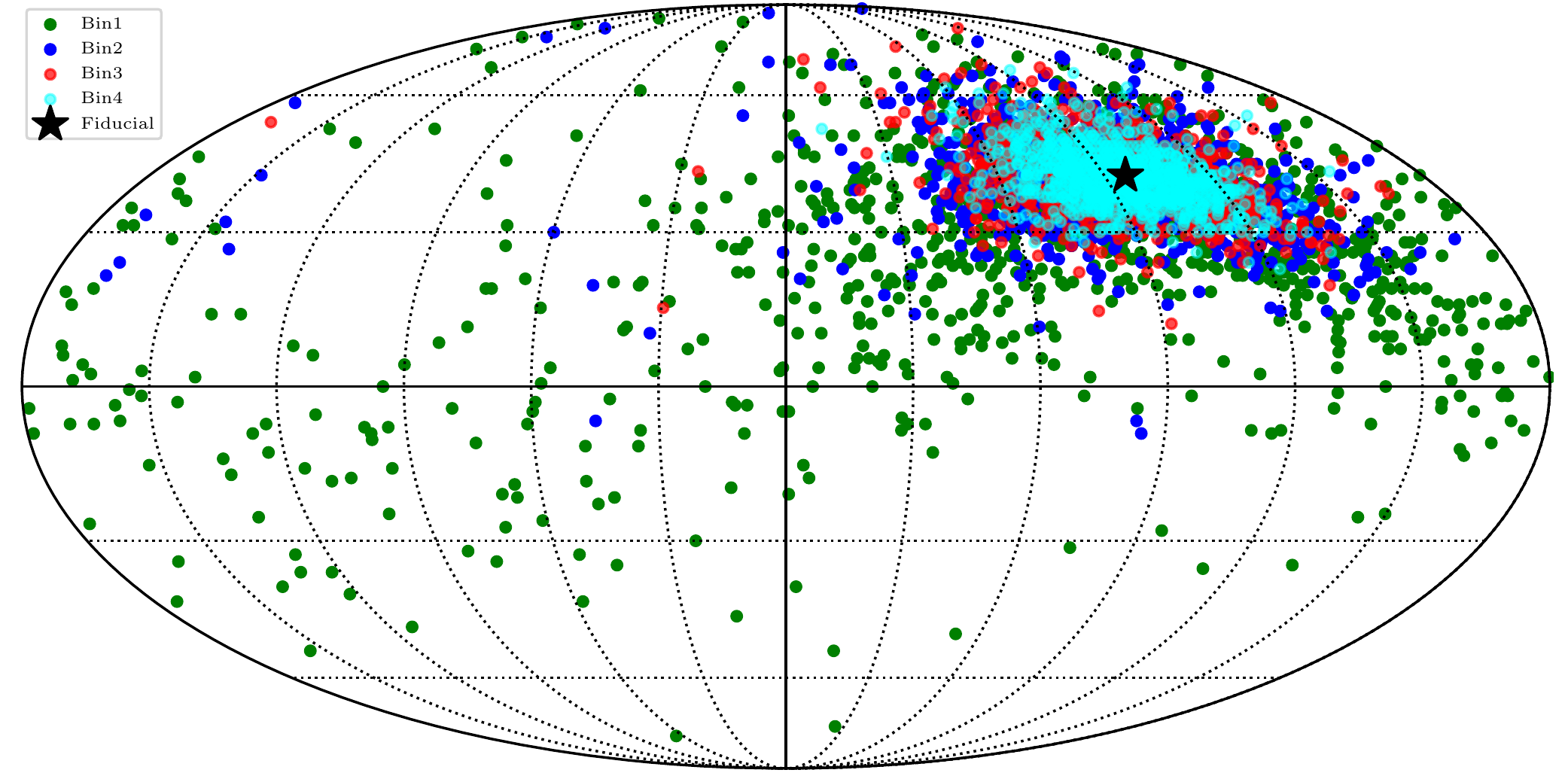}
  \caption{The reconstructed dipole directions with different redshift bins. The green, blue, red and cyan dots indicate results from the first to the fourth redshift bin, respectively. The black star is the fiducial value.}
  \label{fig:zbinang1}
\end{figure}

\begin{figure}
  \centering
  \includegraphics[width=1.0\linewidth]{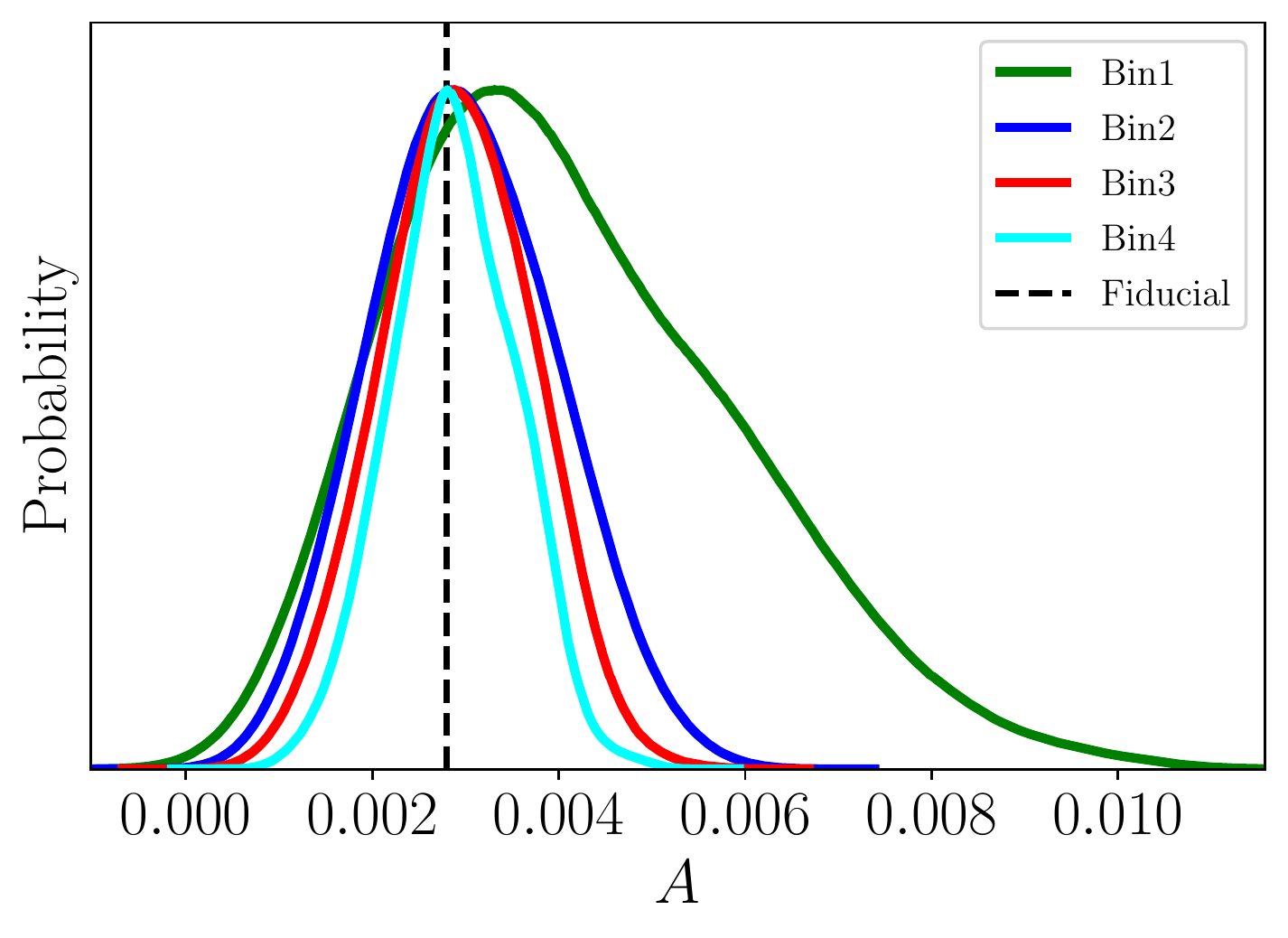}
  \caption{The one-dimensional distributions of reconstructed dipole amplitudes with different redshift bins. The green, blue, red and cyan lines indicate the results from the first to the fourth redshift bin, respectively. The black dotted line is the fiducial value.}
  \label{fig:zbinang2}
\end{figure}

At low redshifts, the kinematic dipole is contaminated by the dipole induced by the nonlinear influence of local large scale structures, and this effect decays with redshift. Since the CSST photometric survey can measure the precise photometric redshifts, we can use different redshift bins to reconstruct the kinematic dipole. In this subsection, we use the four  tomographic bins discussed in Sec.\ref{sec:md}. Firstly, we simulate 1000 mock maps using their selection functions (Eq.(\ref{eq:dndz})) and include the kinematic dipole through a dipole modulation of the number counts (Eq.(\ref{eq:Nobs})). Then we can reconstruct the dipole signal using the quadratic estimator as the last section did. We show the reconstruction results from different redshift bins in Figures \ref{fig:zbinang1} and \ref{fig:zbinang2}. The corresponding quantitative results are listed in Table \ref{tab:zbin}.

\begin{table}
 \caption{The reconstructed dipole directions and amplitudes with different redshift bins.}
 \begin{tabular}{c c c c}
    \hline
    \hline
    Redshift bins  & $b(^{\circ})$ & $l(^{\circ})$ & $A$($\times 10^{-3}$) \\
    \hline
    Bin1 & $65.6\pm20.5$ & $241.1\pm81.1$ & $4.08 \pm 1.86$\\
    Bin2 &  $51.6\pm9.2$ & $263.8\pm35.6$ & $2.99 \pm 1.01$\\
    Bin3 &  $49.9\pm7.2$ & $265.7\pm24.1$ & $2.91 \pm 0.82$\\
    Bin4 &  $49.2\pm5.9$ & $265.5\pm17.2$ & $2.87 \pm 0.67$\\
    \hline
    Fiducial value  & 48.0 & 264.0 & 2.8\\
    \hline
 \end{tabular}
 \centering
 \label{tab:zbin}
\end{table}

We start with the first redshift bin whose samples span between $z=0 \sim 0.6$. The reconstructed dipole direction is $(l,b) = (241.1\pm81.1^{\circ}, 65.6\pm20.5^{\circ})$. The errors are significantly greater than those from the full redshift samples, and the distribution is more dispersed, as shown in Figure \ref{fig:zbinang1}. Besides, the reconstructed dipole amplitude, which is $A = 0.00408 \pm 0.00186$, is weaker than the result from the full redshift samples. As we say, the local structure dipole at low redshifts is a serious contaminant to the cosmic kinematic signal.

Then we use the second redshift bin ($z=0.6 \sim 1.2$) to reconstruct the kinematic dipole signal, and the reconstruction results are $(l,b) = (263.8\pm35.6^{\circ}, 51.6\pm9.2^{\circ})$ and $A = 0.00299 \pm 0.00101$. The errors are smaller than the results from the first bin since we use higher redshift samples, but due to the lack of samples at $z>1.2$, the  reconstruction results are still slightly weaker than the results from full samples.

Finally, the reconstruction results obtained from the third redshift bin ($z=1.2 \sim 1.8$) are $(l,b) = (265.7\pm24.1^{\circ}, 49.9\pm7.2^{\circ})$, $A = 0.00291\pm 0.00082$, and the results are $(l,b) = (265.5\pm17.2^{\circ}, 49.2\pm5.9^{\circ})$, $A = 0.00287 \pm 0.00067$ using the fourth redshift bin ($z=1.8 \sim 4$). The reconstructions from the third and fourth redshift bins are stronger than the results from the full redshift samples. We can see, at $z=0 \sim 4$, higher redshift samples are less affected by the local structure dipole. Using the fourth redshift bin, we can detect the kinematic dipole signal at 4$\sigma$ confidence level. Our errors are slightly larger than the reconstruction results using the future SKA \citep{Bengaly:2018ykb}. The reason is the  SKA can detect higher redshift samples and the dipole amplitude $A$ of radio sources are larger than the  optical sources. Nevertheless, we can still use the results from optical measurements as a consistency check.

It should be emphasized that although we can obtain better reconstruction results with higher redshift samples, we cannot use a narrow range at very high redshifts, such as $z = 3.5 \sim 4$. The reason is there are fewer samples, which may lead to large variations in the number density of sources across the sky.

\subsection{Signal to noise estimate}

Finally, we can make an estimate of the signal-to-noise ratio (SNR) for the kinematic dipole detection, using the expression in \citet{Baleisis:1997wx, Itoh:2009vc, Bengaly:2017slg}
\be
\label{eq:SNR}
\mathrm{SNR}=\frac{A_{\mathrm{kin}}}{\sqrt{A_{\mathrm{LSS}}^{2}+A_{\mathrm{PN}}^{2}}},
\ee
where $A_{\rm kin}$, $A_{\rm LSS}$, $A_{\rm PN}$ are the dipole amplitudes from kinematic effect, large-scale structure and Poisson noise, respectively. Here $A_{\rm kin}$ is given by the mean value obtained from the quadratic estimator (listed in Tables \ref{tab:zall} and \ref{tab:zbin}). $A_{\rm LSS}$ is obtained using the same estimator but we input mock maps without dipole modulation.
$A_{\rm PN}$ is given by \citep{Bengaly:2017slg}:
\be
A_{\mathrm{PN}}=\frac{3}{2} \sqrt{\frac{4 f_{\mathrm{sky}}}{N_{\rm total}}}.
\ee
There are 477, 695, 278, 224 million galaxies in the four redshift bins, respectively, so the contributions of $A_{\rm PN}$ is puny.

In Table \ref{tab:snr}, we list $A_{\rm kin}$, $A_{\rm LSS}$, $A_{\rm PN}$ and SNR estimated from the four redshift bins and the full redshift. We can see the SNR increases with redshift, and the last bin gives $\rm SNR = 3.63$. This result indicates that the CSST photometric survey can help us extract the kinematic dipole signal effectively.

\begin{table}
	\caption{$A_{\rm kin}$, $A_{\rm LSS}$, $A_{\rm PN}$ and SNR estimated from the four redshift bins and the full redshift.}
    \setlength{\tabcolsep}{1mm}{
	\begin{tabular}{c c c c c}
    \hline
    \hline
    Redshift bins & $A_{\rm kin}(\times 10^{-3})$ & $A_{\rm LSS}(\times 10^{-3})$ & $A_{\rm PN}(\times 10^{-3})$ & SNR \\
    \hline
    Bin1 &  4.08 & 3.17 & 0.087 & 1.29 \\
    Bin2 &  2.99 & 1.29 & 0.072 & 2.31 \\
    Bin3 &  2.91 & 1.01 & 0.114 & 2.86 \\
    Bin4 &  2.87 & 0.78 & 0.127 & 3.63 \\
    Total&  2.96 & 1.25 & 0.046 & 2.37 \\
    \hline
	\end{tabular}}
	\centering
	\label{tab:snr}
\end{table}

\section{Discussions and Conclusions}
\label{sec:4}

The cosmological principle is a fundamental hypothesis of the standard model of cosmology, and it is crucial to take the validity test. Due to the relative motion between the Solar System and the cosmic rest-frame, we can probe the consistency between the kinematic dipoles measured in the CMB and the large scale structure. A mismatch between these dipoles may indicate a violation of the cosmological principle or a sign of new features on the large scales.

In this paper, we investigate the capacity of the future CSST photometric survey to reconstruct the dipole direction and amplitude by simulating CSST number count maps. We find that the CSST photometric survey can detect the kinematic dipole signal at $3\sigma$ C.L. and $(\Delta l,  \Delta b) \sim (30.8^{\circ},    8.5^{\circ})$ on the direction using the full redshift samples.

Since the kinematic dipole at low redshifts is contaminated by the local structures, we  reassess our analysis with different redshift bins. We find the reconstruction results get better with higher redshift samples. Using the samples span between $z=1.8 \sim 4$, we can detect the kinematic dipole signal at $4\sigma$ C.L., and the standard deviation of direction is $(\Delta l, \Delta b) \sim (17.2^{\circ}, 5.9^{\circ})$. We also estimate the SNR for the kinematic dipole detection, and we obtain $\rm SNR = 3.63$ using the samples at $z=1.8 \sim 4$.

In our analysis, we assume $p=0$ and $x=0.11$ as \citet{Yoon:2015lta} did, which leads to $A=0.0028$. \citet{Yoon:2015lta} pointed the standard cosmology theory predicts  the actual value of $A$ takes our fiducial value, plus or minus O(20\%) changes depending on the source population selected, and it gives $A_{\rm actual} = 0.00224\sim0.00336$. Based on the possible values of  $A$, we get $\sigma(A)/A = 0.184 \sim 0.265$ using  the  fourth redshift bin and the quadratic estimator, which means we can detect the kinematic dipole signal at $ 3.77\sigma \sim 5.43\sigma$ confidence level. We can also calculate the SNR using Eq.(\ref{eq:SNR}), and the results is ${\rm SNR} = 2.97 \sim 4.33$.

Finally, we must emphasize that the galaxy bias may affect the reconstruction results. As a check, we calculate another case with halo mass $M = 10^{13}~h^{-1}M_{\odot}$. The resulting bias is 1.5-2 times larger than the bias used in our work. Using the full redshift samples and the quadratic estimator, we get $(l,b) = (256.2\pm50.2^{\circ}, 55.6\pm13.2^{\circ})$ and $A = 0.00329 \pm 0.00127$. Clearly, a larger bias can make it harder to detect the kinematic dipole signal. Due to the bias factor $b_{\rm g}(z)$ increases with redshifts, the reconstruction accuracy of the dipole cannot always improve with the increase of redshifts. Besides, since we do not know the detailed evolution of  bias, we also redo our analysis with a constant bias $b(z)=1.5$. This constant bias is larger than the halo bias used in our work at $z<1.25$, which leads to larger reconstruction errors using the first two bins. On the contrary, the third and the fourth bin can give tighter reconstructions using the constant bias. The results show $\sigma(A)|_{\rm constant}/\sigma(A)|_{\rm halo} = 1.46, 1.15, 0.87, 0.64$ for the first bin to the last bin, respectively. The last bin gives the best reconstruction: $(l,b) = (263.6\pm11.4^{\circ}, 48.7\pm3.9^{\circ})$ and $A = 0.00286 \pm 0.00043$.
Since kinematic dipole is contaminated by the low redshift samples, the reconstructed results using constant bias are weaker when we consider the full redshift samples, which are $(l,b) = (267.1\pm48.6^{\circ}, 55.4\pm12.9^{\circ})$ and $A = 0.00327 \pm 0.00122$. 

\section{DATA AVAILABILITY}
The mock number count maps of CSST will be shared on reasonable request to the corresponding author.

\section*{Acknowledgements}
% Entry for the table of contents, for this guide only
J.-Q. Xia is supported by the National Science Foundation of China, under grant Nos. U1931202 and 12021003,
and the National Key Research and Development Program of China, No. 2017YFA0402600.

% Don't change these lines
\bsp	% typesetting comment
\label{lastpage}

\end{document}